\documentclass{kapproc} 

\RequirePackage{graphicx}%
\RequirePackage{epsf}%
\input{psfig.sty}

\upperandlowercase
\let\footnote\savefootnote
\let\footnotetext\savefootnotetext 
 
\kluwerbib 

\begin{document}


\articletitle{Tiny HI Clouds in the Local ISM}


\chaptitlerunninghead{Tiny Atomic Clouds}





\author{Robert Braun\altaffilmark{1} and Nissim Kanekar\altaffilmark{2}}

\affil{\altaffilmark{1}ASTRON, Postbox 2, 7990 AA Dwingeloo, The Netherlands,
 \\ 
\altaffilmark{2}Kapteyn Inst., Postbox 800, 9700 AV Groningen, The Netherlands}


 \begin{abstract}
Very sensitive HI absorption spectra ($\Delta\tau\sim10^{-4}$ over 1~km/s)
toward high latitude QSOs have revealed a population of tiny discrete features
in the diffuse ISM with peak $\tau$ of 0.1 -- 2\% and core line-widths
corresponding to temperatures as low as 20~K. Imaging detections confirm
linear dimensions of a few 1000~AU. We suggest these structures may be formed
by the stellar winds of intermediate mass stars. A more speculative origin
might involve molecular ``dark matter''.
 \end{abstract}

\section{Introduction}

Over the past decades there have been several lines of evidence suggesting a
surprising degree of small-scale structure in the atomic interstellar medium.
One of the first of these was the observation of spatially variable HI
absorption seen toward compact radio sources (eg. Dieter et al. \cite{diet76},
Davis et al. \cite{davi96}, Faison et al. \cite{fais98}, Faison \& Goss
\cite{fais01}). In the most extreme cases, for example toward 3C138, there is
good evidence for a variation in the HI opacity, $\tau~=~1.7(n \cdot s)/(T
\Delta V) [cm^{-3}pc/K-km/s]$, of as much as $\Delta\tau$~=~0.1 on 20~AU
transverse scales (Faison et al. \cite{fais98}). The ``simplest''
interpretation of these observations are very large variations in the volume
density, $\Delta$n$_{HI}~\sim~10^5$~cm$^{-3}$, assuming that all of the other
relevant variables (specifically pathlength, s, and temperature, T) are kept
fixed.  However, as argued by Deshpande (\cite{desh00}), realistic ISM
structure functions can lead to large variations of $\tau$ with small
angular offsets simply from the statistical fluctuations in the effective
pathlength with position. Another line of evidence for small-scale atomic
structure came from searches for time variability in the HI absorption seen
toward pulsars (eg. Frail et al. \cite{frai94}, Johnston et al. \cite{john03},
Stanimirovic et al. \cite{stan03}). However, the early claims for ubiquitous
and significant time variations in $\tau$ have not been confirmed in the
recent careful studies. A third line of evidence has come from observations
of Na~I observation toward nearby pairs of stars (Watson \& Meyer
\cite{wats96}, Lauroesch et al. \cite{laur98}). These observations have
revealed strong evidence for discrete absorption features in the local ISM
which have highly variable properties on 100's of AU scales. Since discrete
features are being probed by these observations, rather than simply the
projection along a long line-of-sight, the structure function arguments of
Deshpande do not seem to apply.

\section{Observations}

We have undertaken a series of extremely sensitive HI absorption observations
toward bright background QSOs near the North Galactic Pole (NGP) utilizing the
Westerbork Synthesis Radio Telescope (WSRT). The initial motivation for these
observations was detection of HI absorption from a Warm Neutral Medium (WNM),
even if it's temperature was as high as 10$^4$~K. Such an experiment was
prompted by the detection of HI absorption by Kanekar et al. (\cite{kane03})
at velocity widths extending at least as high as an equivalent temperature of
3500~K, which corresponded to the sensitivity limit of those data. The NGP
region was chosen since it was expected that the lines-of-sight through our
Galaxy disk might be as short as possible and therefore relatively
simple. Sensitive data were acquired toward 3C286 (14.7~Jy), 3C287 (7~Jy),
4C+32.44 (5~Jy) and B2~1325+32 (1.4~Jy) by observing in an in-band
frequency-switching mode utilizing a 1~MHz throw every 5 minutes inside a
2.5~MHz total bandwidth with 0.5~km/s channel width. The in-band frequency
switching allowed exceptionally good band-pass calibration while providing
100\% of the observing time on-source. For the brightest sources we achieve an
RMS $\Delta\tau~<~2\times10^{-4}$ over only 1~km/s, making these the most
sensitive HI (21cm) absorption measurements of which we are aware.

\section{Results}

\begin{figure}[ht]
\centerline{\psfig{file=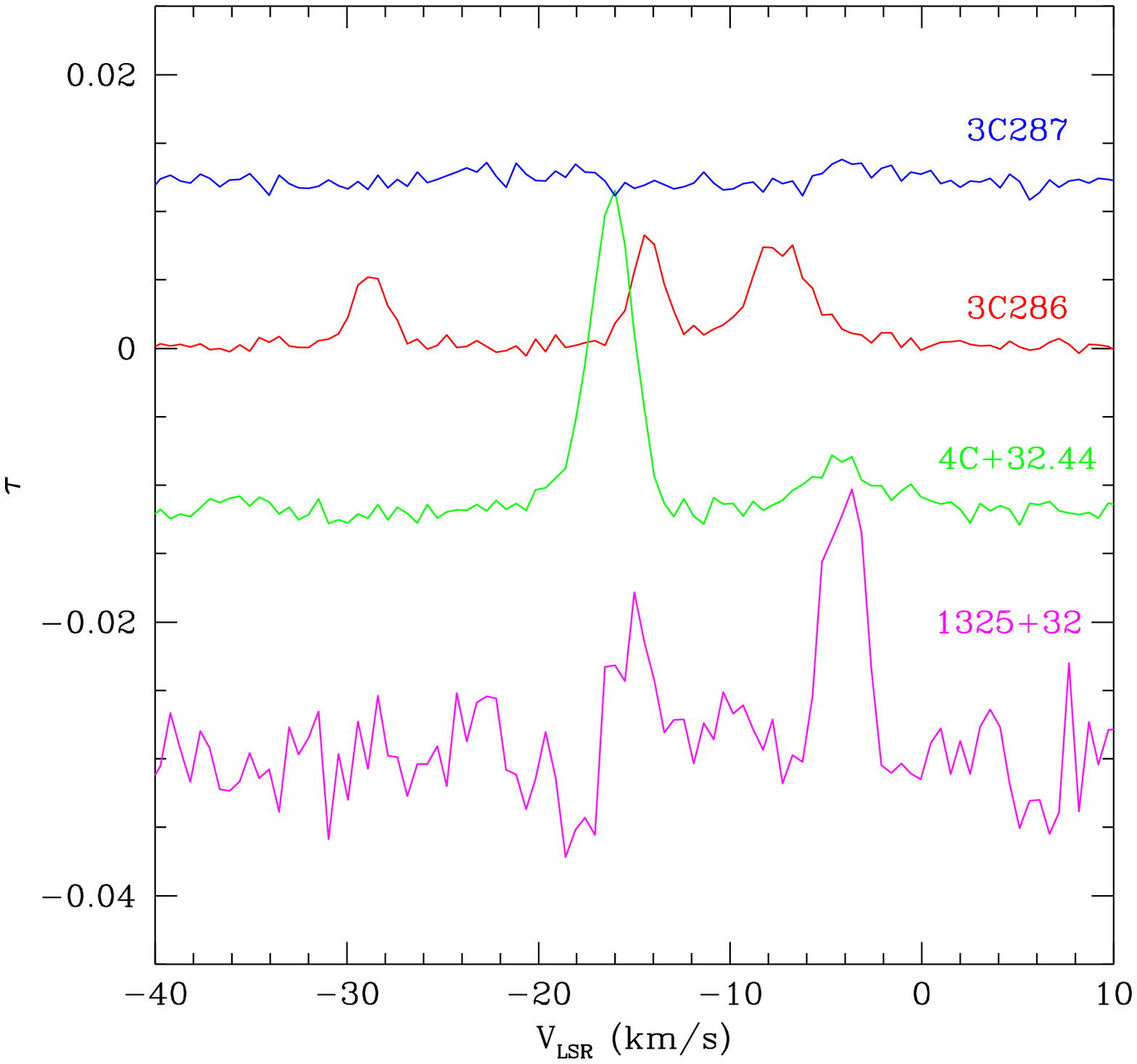,width=5cm},
  \psfig{file=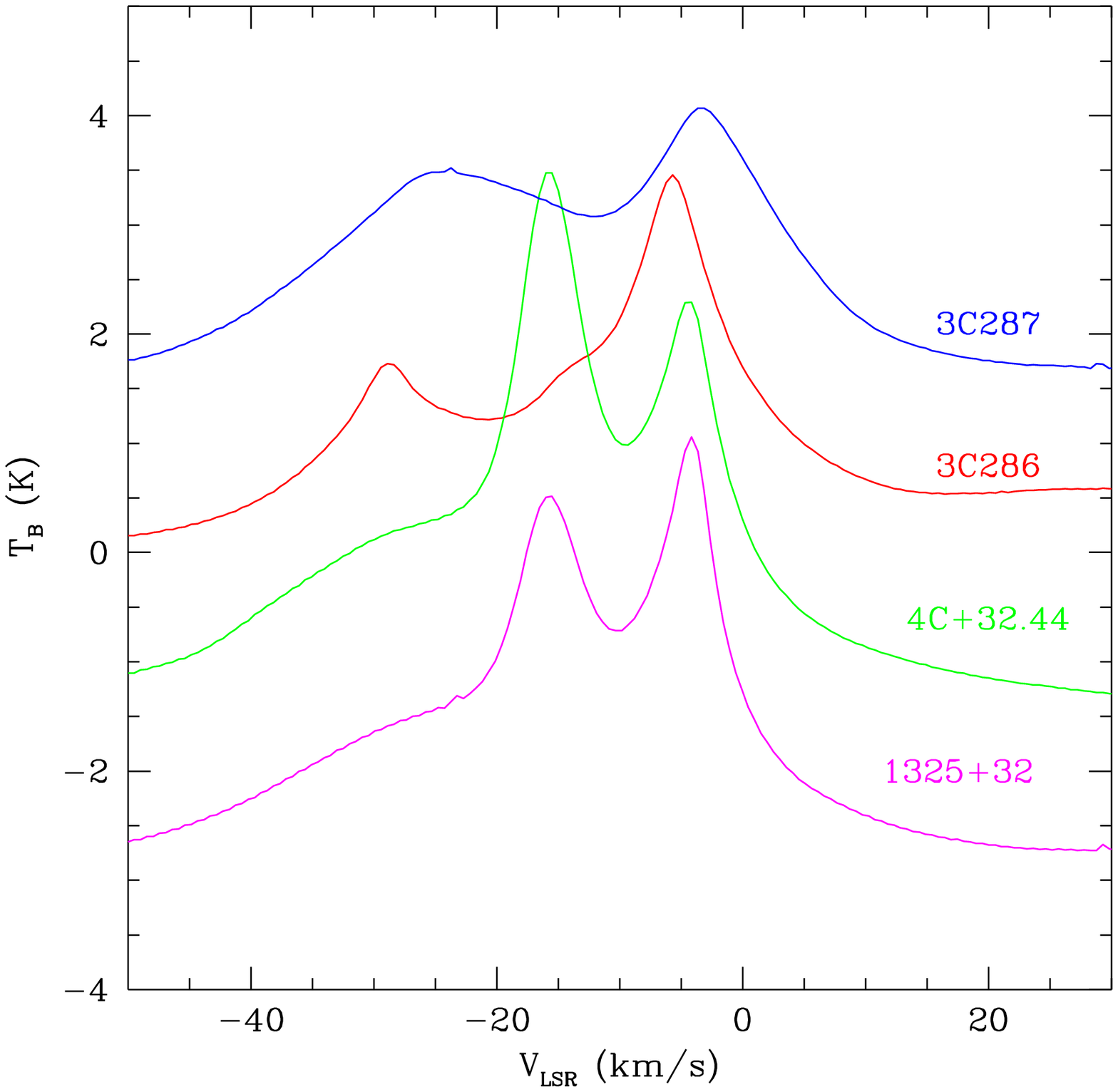,width=5cm}} 
\caption{HI absorption (left) and total power emission (right) spectra.}
\label{fig:allspec}
\end{figure}

Rather than detecting only the simple, broad and shallow absorption profile of
the WNM toward these high latitude lines-of-sight, we were surprised to find
instead that multiple narrow absorption features were detected at discrete
line-of-sight velocities with peak opacities in the range 0.1 to 2\%. The
detection of discrete absorption features was particularly surprising since
the peak emission brightness seen in these directions (with the 35~arcmin
total power beam) was only between about 2 and 5~K.  The four observed
lines-of-sight are compared in Figure~\ref{fig:allspec}. Since many of the
absorption lines are detected with high signal-to-noise it is possible to say
with confidence that the profiles of individual features are non-Gaussian, and
instead appear to be semi-Lorentzian with very narrow line-cores that merge
smoothly into broader wings. Of course it is always possible to represent each
feature with a sum of coincident Gaussians, but such a procedure appears to be
quite arbitrary and it is far from clear what physical interpretation might be
attached to such an arbitrary decomposition.

Moderately near-by lines-of-sight show essentially uncorrelated absorption
features and only weakly correlated total power emission features. Only in the
case of 4C+32.44 and B2~1325+32 which have an angular separation of only
15~arcmin is there some possibility of similar velocity components having been
detected.

\section{Analysis}

Motivated by the non-Gaussian line profiles we have explored some simple
spherically symmetric, iso-baric cloud models of the form:
\begin{equation}
n_H(r) = n_o exp[-(r/s)^{\alpha_1}] \hskip1cm {\rm for~T~<~4000~K~and} 
\end{equation}
\begin{equation}
n_H(r) = n_o exp[-(r/s)^{\alpha_2}] \hskip1cm {\rm for~T~>~4000~K~~~~\ }
\end{equation}
where we further relate volume density to temperature using the thermal
pressure which was assumed to be constant at $P/k_B~=~n_HT$ =
1500~cm$^{-3}$K. The temperature was allowed to vary between T$_{min} = 20~K$
and T$_{max} = 15000~K$, yielding an assumed thermal velocity dispersion of
$\sigma^2~=~0.0086 T$. The predicted HI absorption and total power emission
spectra were calculated for a ``cloud'' placed at the central velocity of each
observed feature in an attempt to simultaneously reproduce the observed
spectra shown in Fig.~\ref{fig:allspec}. The most important free parameters in
this process were the cloud scale-length, $s$, the cloud distance, $d$ (which
most strongly influences the predicted total power emission) and an impact
parameter, $b$. This last parameter was used to allow for the likely
circumstance that each spherical model cloud may not be penetrated exactly
on-axis by the background absorber.  The two power-law indices of the scaled
distance, $\alpha_1$ and $\alpha_2$ determine the characteristic line-shape in
the cold core and warm halo respectively. Although these were in principle
also free parameters, it was found that only minor variations from
``standard'' values of about $\alpha_1$=1/4 and $\alpha_2$=1/8 were needed.

\begin{figure}[ht]
\centerline{\psfig{file=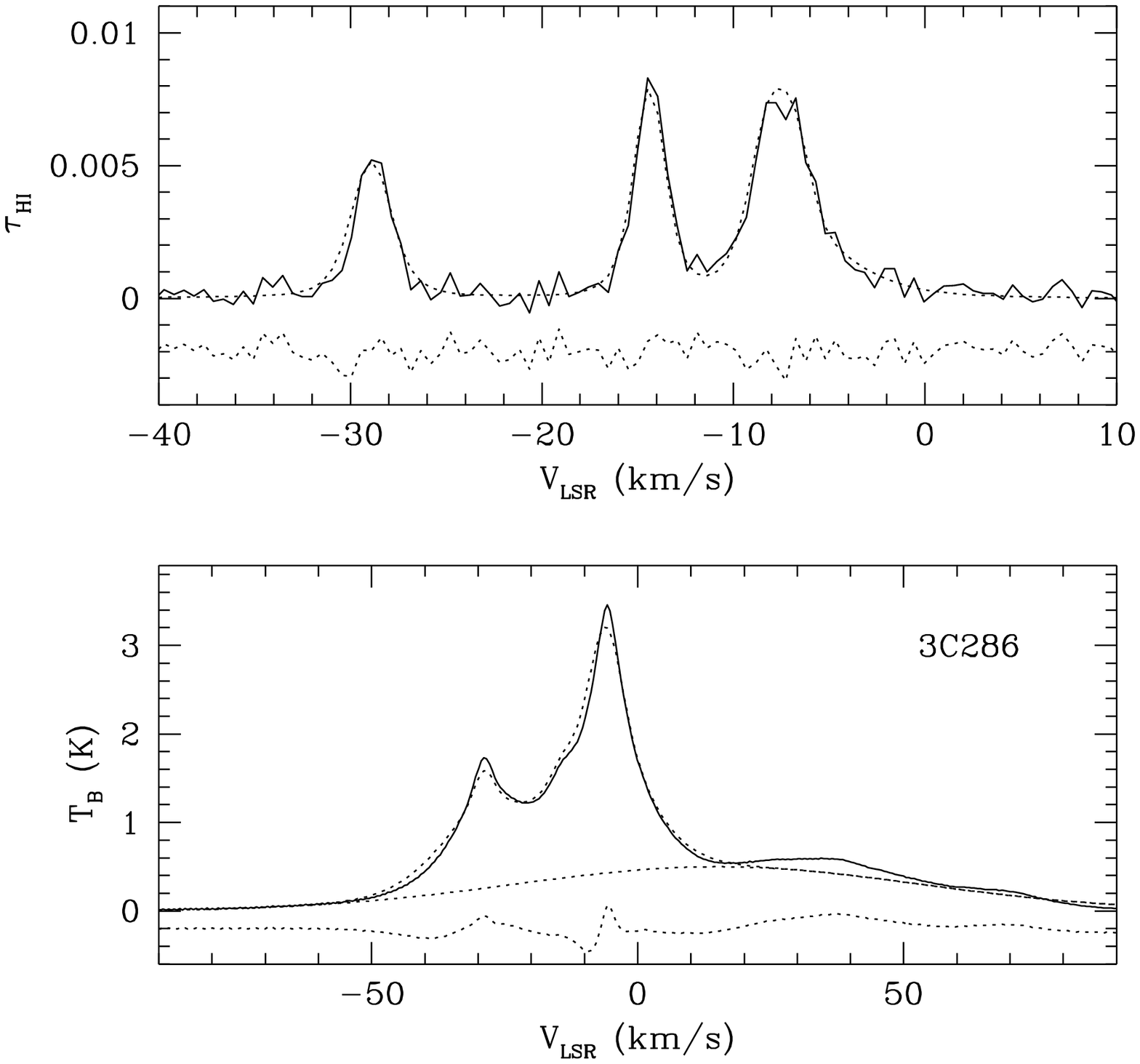,width=5cm},
  \psfig{file=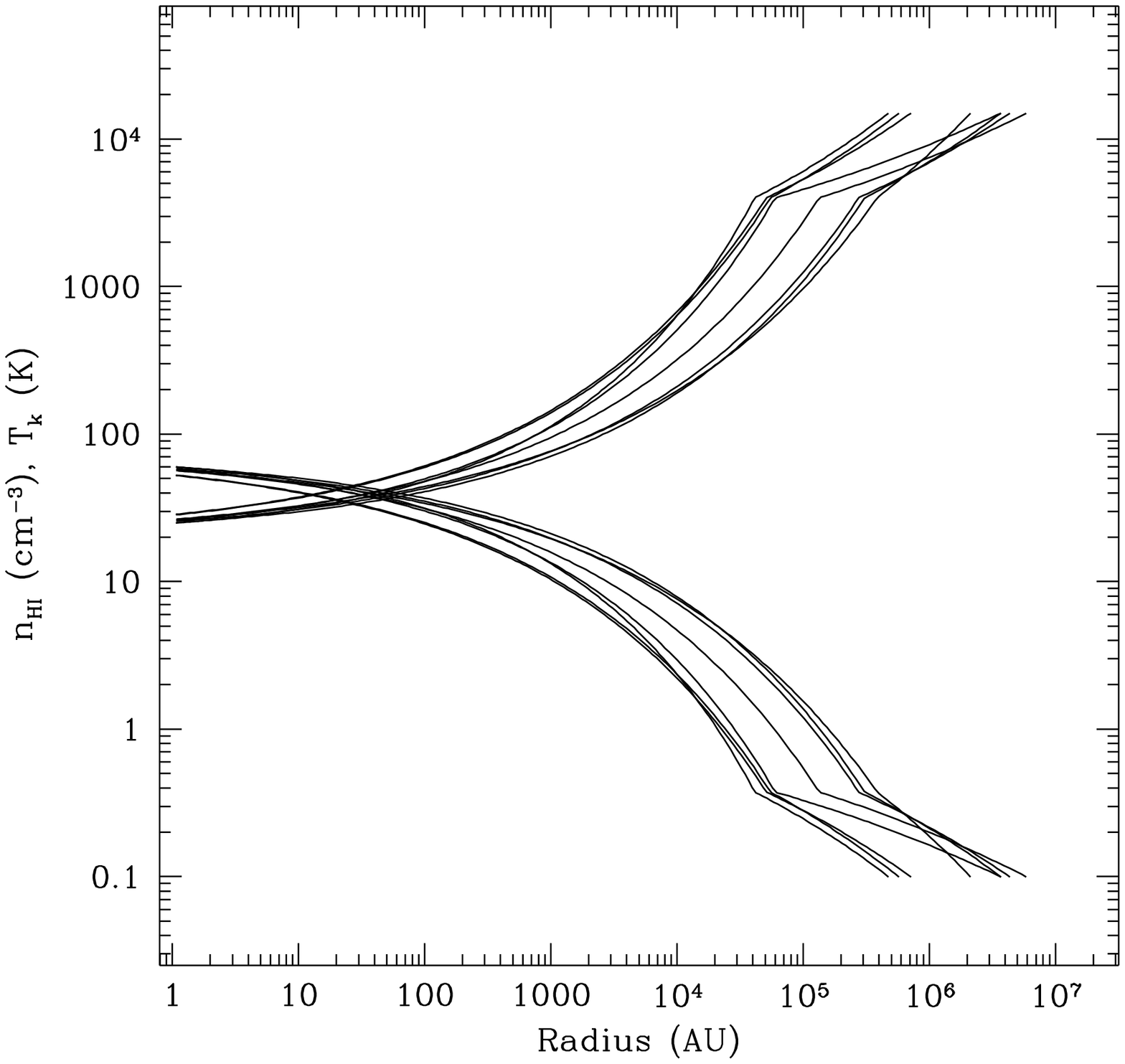,width=5cm}} 
\caption{Overlay of observed and modelled spectra for the 3C286 l-o-s
  (left). Density and temperature profiles of all modelled ``clouds'' (right).}
\label{fig:modcomp}
\end{figure}

An illustration of the simultaneous emission and absorption line fitting is
shown in Fig.~\ref{fig:modcomp} for the 3C286 line-of-sight. No $\chi^2$
fitting has actually been carried out, but merely a $\chi$-by-eye to
illustrate the possibilities of this approach. The entire set of density and
temperature profiles for the required clouds along all four lines-of-sight is
also shown in the Figure. This illustrates the basic similarities of the
modelled features, although the apparent scale-lengths do vary by about an
order of magnitude. The apparent distances used in reproducing the spectra was
about 10~pc, and the typical half-density radius was 100~AU, although these
were not very well-constrained given the very low angular resolution of our
total power data (35~arcmin). These modelling results can be easily scaled to
other assumed thermal pressures by a linear scaling of $n_H$ together with an
inverse linear scaling of both $s$ and $d$. For example, with a typical thermal
pressure of only 150~cm$^{-3}$K, comparable fits would be obtained at apparent
distances of 100~pc and typical half-density radii of 1000~AU. Variations from
the idealized spherical cloud symmetry would have a similar impact on apparent
distances and sizes.
 
\begin{figure}[ht]
\centerline{\psfig{file=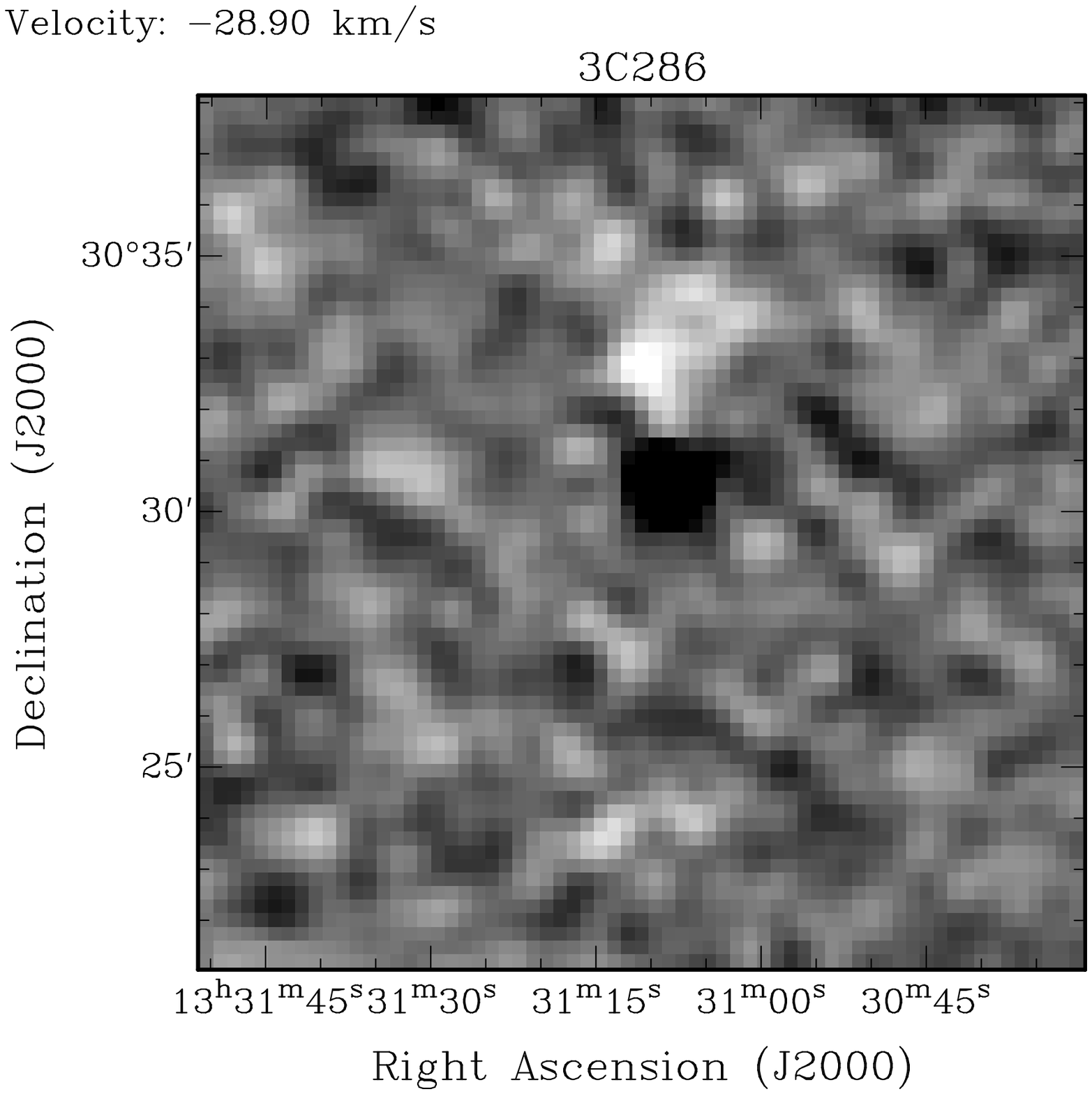,width=5cm},
  \psfig{file=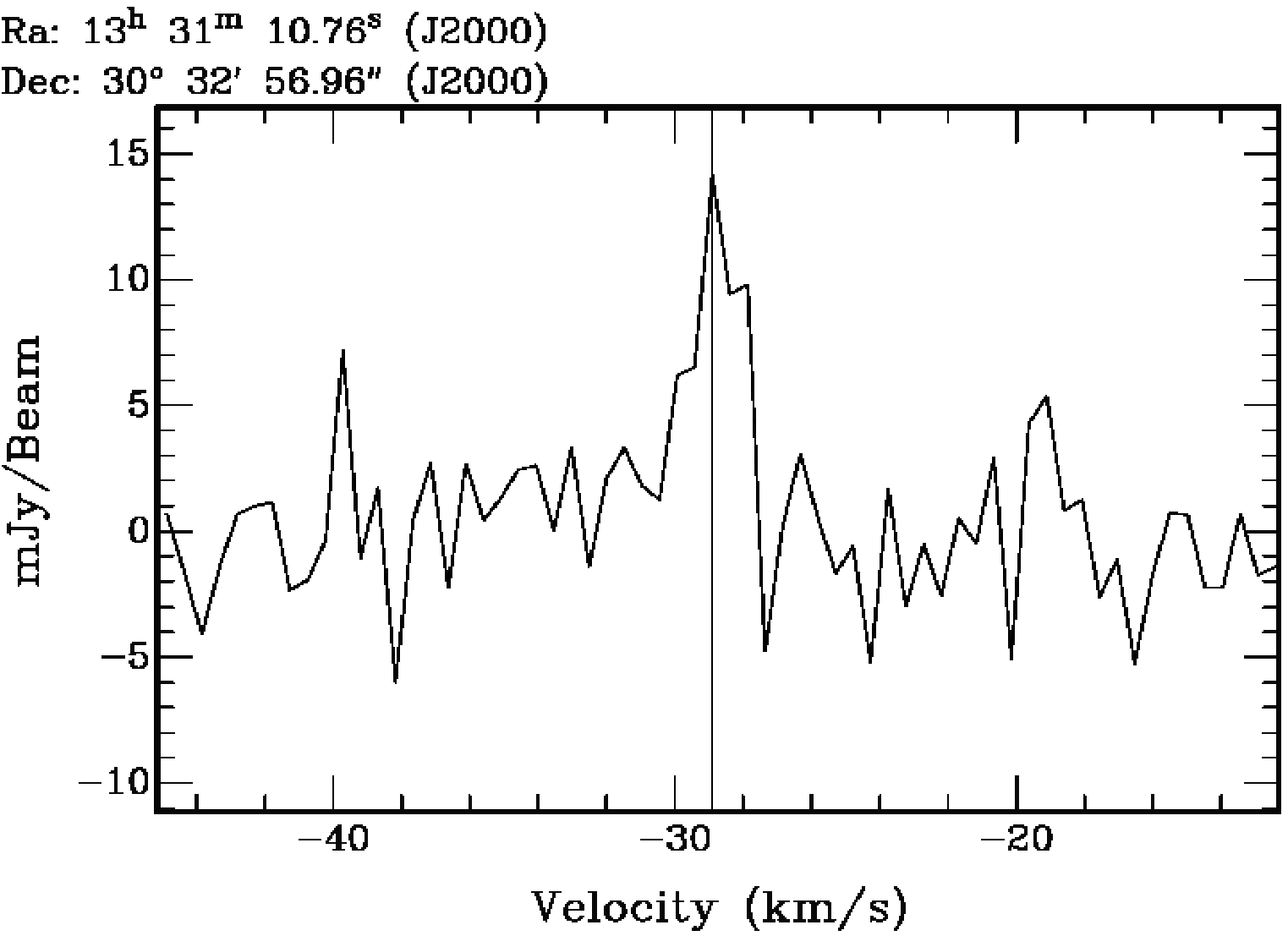,width=5cm}} 
\caption{HI emission clump adjacent to 3C286 in map (left) and
  spectrum (right).}
\label{fig:3clump}
\end{figure}

We also have more concrete information regarding the nature of these features,
since the same WSRT observations that provided the absorption spectra also
allow an imaging search for emission counterparts. Cubes of HI line emission
were produced at a range of angular resolutions (15, 30, 60 and 120
arcsec). Obtaining sufficient brightness sensitivity to achieve detections in
emission typically required smoothing to 60~arcsec. At those velocities where
total power emission exceeding about 1~K is seen in Fig.~\ref{fig:allspec}
compact emission clumps of 2--3 K brightness are detected at apparently random
locations in the field superposed on the poorly-sampled diffuse background
emission. One such emission clump, immediately adjacent to the 3C286
line-of-sight is shown in Fig.~\ref{fig:3clump}. The intrinsic angular size of
these clumps appears to be about 30~arcsec, while their FWHM line-widths are
1--2~km/s, corresponding to the thermal linewidths of 20--80~K HI. A
representative detected column density for the clumps in the 3C286 field is
N$_{HI}~\sim~5\times10^{18}$cm$^{-2}$. At an assumed distance of say, 100~pc,
the clump size would correspond to 3000~AU and the central volume density
would be about 100~cm$^{-3}$.

\begin{figure}[ht]
\centerline{\psfig{file=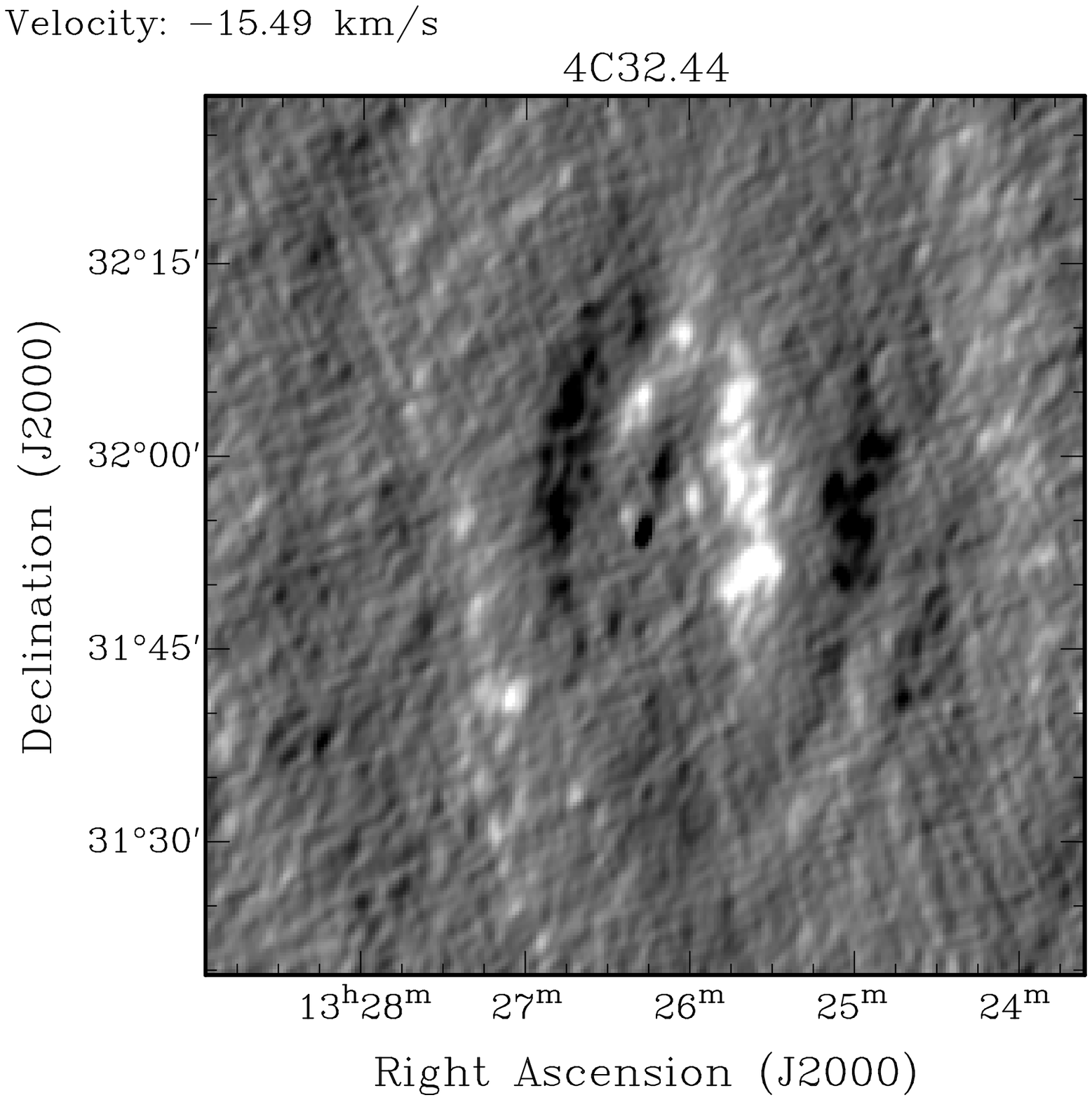,width=4.1cm},
  \psfig{file=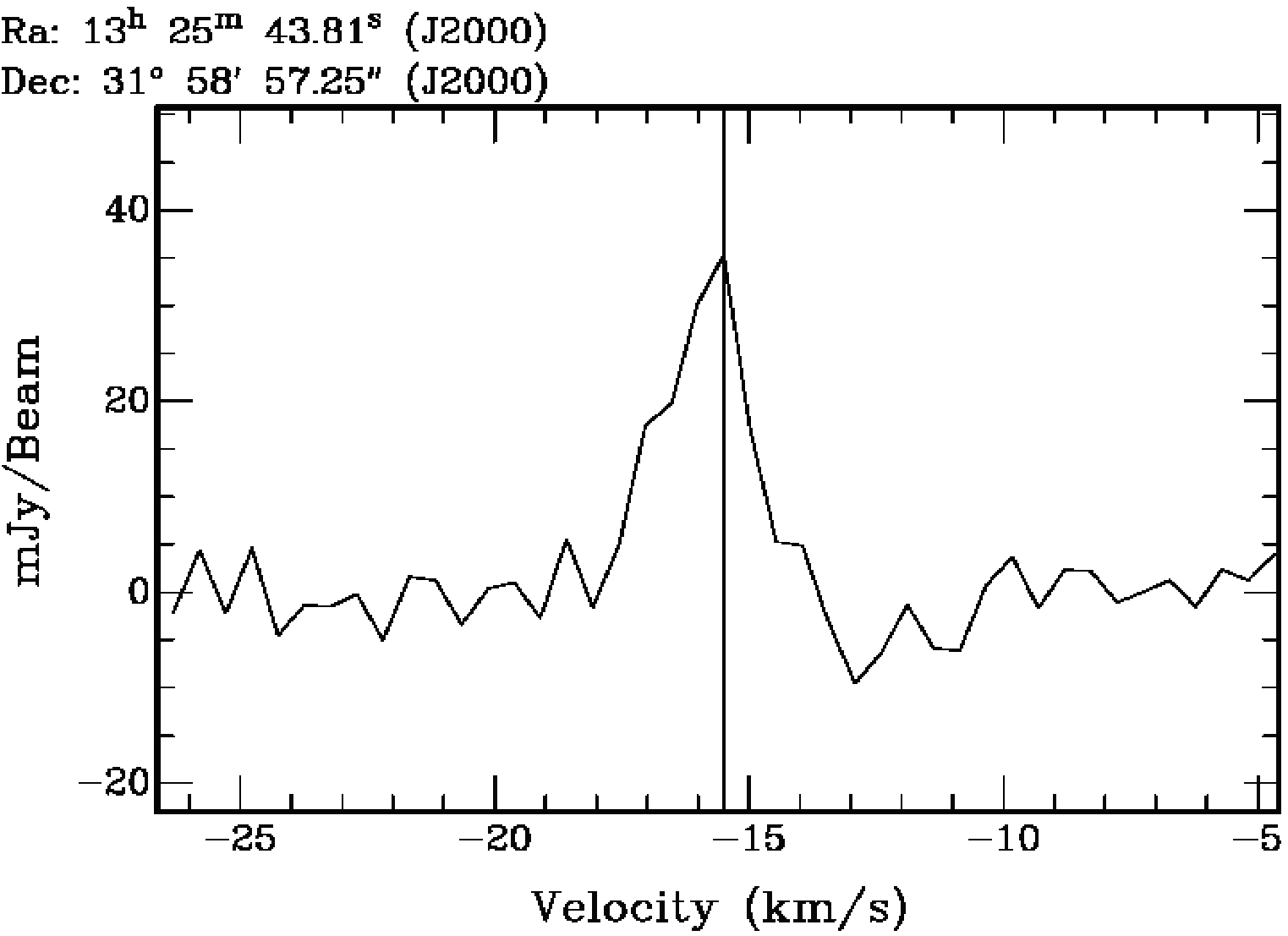,width=5cm}} 
\caption{HI emission shell toward 4C+32.44 in map (left) and
  spectrum (right).}
\label{fig:4shell}
\end{figure}

A more interesting emission structure is detected in the 4C+32.44 field. The
line-of-sight toward this background source appears to intersect a 15~arcmin
diameter shell of HI emission as shown in Fig.\ref{fig:4shell}. Although it
may be a chance super-position, this apparent shell includes the G0III star
HD~116856 at $(\alpha,\delta)_{2000}$ = (13:25:55.835,+31:51:40.629). The
measured parallax of this star places it at 105$\pm$11~pc, where the shell
would have a 0.45~pc diameter. The stellar proper motion
$(\Delta\alpha,\Delta\delta)$ = (+14.73,$-$43.40) mas/yr is directed
predominatly toward the South. Peak column densities in this structure reach
N$_{HI}~>~10^{19}$cm$^{-2}$ with FWHM line-widths of 2--3~km/s.

\section{Discussion}

Although more work needs to be done to fully characterize the type and
quantity of sub-structure in the ``diffuse'' ISM, it is already becoming clear
that even the most diffuse regions are populated by tiny distinct structures
of very high density- and temperature-contrast. For example, the low
scale-size end of the turbulent power spectrum is predicted (Deshpande
\cite{desh00}) to have HI opacity fluctuations of only 10$^{-5}$ on spatial
scales of 1000~AU. In fact, we measure more than two orders of magnitude
larger opacity fluctuations of $>10^{-3}$ on these scales. There appears to be
substantial injection of fluctuation power on very small scales. The physical
origin of these tiny structures is not yet clear, but it seems conceivable
that the stellar winds of intermediate mass stars may play an important role
in their formation, whenever such stars find themselves within a diffuse
atomic structure. A more speculative origin might be some relation to
molecular ``dark matter'' (eg. Pfenniger \& Combes \cite{pfen94}).
New observations should test various scenarios.

%

\end{document}